# SPARCQ: A new approach for fat fraction mapping using asymmetries in the phase-cycled bSSFP signal profile

Giulia MC Rossi[1,2,3], Adèle LC Mackowiak[1,2,3], Tom Hilbert[3,4,5], Berk Can Açikgöz[1,2], Katarzyna Pierzchała[6,7,8], Tobias Kober[3,4,5], Jessica AM Bastiaansen[1,2]

[1]Department of Diagnostic, Interventional and Pediatric Radiology (DIPR), Inselspital, Bern University Hospital, University of Bern, Switzerland

[2]Translational Imaging Center, Swiss Institute for Translational and Entrepreneurial Medicine, Bern, Switzerland

[3]Department of Diagnostic and Interventional Radiology, Lausanne University Hospital and University of Lausanne, Lausanne, Switzerland

[4]Advanced Clinical Imaging Technology, Siemens Healthineers International AG, Lausanne, Switzerland

[5]LTS5, École Polytechnique Fédérale de Lausanne (EPFL), Lausanne, Switzerland

[6]Laboratory of Functional and Metabolic Imaging, Swiss Federal Institute of Technology, Lausanne, Switzerland

[7]CIBM Center for Biomedical Imaging, Switzerland

[8]Animal Imaging and Technology, Ecole Polytechnique Fédérale de Lausanne, Lausanne, Switzerland

**To whom correspondence should be addressed:** Jessica AM Bastiaansen, Department of Diagnostic, Interventional and Pediatric Radiology (DIPR), Inselspital, Bern University Hospital, University of Bern, Switzerland. Freiburgstrasse 3, 3010 Bern, Switzerland,

Email: jbastiaansen.mri@gmail.com

twitter: @jessica_b_

# ABSTRACT

**Purpose:** To develop SPARCQ (Signal Profile Asymmetries for Rapid Compartment Quantification), a novel approach to quantify fat fraction using asymmetries in the phase-cycled bSSFP profile.

**Methods:** SPARCQ uses phase-cycling to obtain bSSFP frequency profiles, which display asymmetries in the presence of fat and water at certain repetition times. For each voxel, the measured signal profile is decomposed into a weighted sum of simulated profiles via multi-compartment dictionary matching. Each dictionary entry represents a single-compartment bSSFP profile with a specific off-resonance frequency and relaxation time ratio. Using the results of dictionary matching, the fractions of the different off-resonance components are extracted for each voxel, generating quantitative maps of water and fat fraction and banding-artifact-free images for the entire image volume. SPARCQ was validated using simulations, experiments in a water-fat phantom and in knees of healthy volunteers. Experimental results were compared with reference proton density fat fractions obtained with $^{1}$H-MRS (phantoms) and with multiecho gradient-echo MRI (phantoms and volunteers). SPARCQ repeatability was evaluated in six scan-rescan experiments.

**Results:** Simulations showed that fat fraction quantification is accurate and robust for signal-to-noise ratios greater than 20. Phantom experiments demonstrated good agreement between SPARCQ and gold standard fat fractions. In volunteers, banding-artifact-free quantitative maps and water-fat-separated images obtained with SPARCQ and ME-GRE demonstrated the expected contrast between fatty and non-fatty tissues. The coefficient of repeatability of SPARCQ fat fraction was 0.0512.

**Conclusion:** SPARCQ demonstrates potential for fat quantification using asymmetries in bSSFP profiles, and may be a promising alternative to conventional fat fraction quantification techniques.

# 1. INTRODUCTION

Accurate fat quantification is crucial for a variety of clinical applications, including the evaluation of liver steatosis[1–3], fatty infiltration in the myocardium which is associated with heart failure[4,5], the relationship between osteoporosis and bone marrow adiposity[6,7], and characterization of cellularity for radiation dosimetry in cancer patients[8]. Noninvasive fat quantification by MRI has improved patient comfort by avoiding invasive biopsies, which are prone to sampling errors and are highly invasive, limiting their usage in obtaining prognostic information[4].

Fat quantification can be achieved by quantitative proton density fat fraction (PDFF) methods such as single-voxel magnetic resonance spectroscopy (MRS), or chemical-shift-based water-fat imaging such as multiecho Dixon MRI[9–14]. Multiecho techniques rely on the acquisition of multiple echoes and require advanced signal processing for robust water fat separation and quantification[12,15–17]. Multi-parametric mapping techniques have been combined with multiecho sampling to enable fat quantification, with promising results[18–21]. Although powerful, PDFF quantification using multiecho methods can be confounded by $T_1$ bias[22,23], the assumption of a single T2* decaying component[24], the use of a predetermined spectral fat model[2,22,25], and the inherent low signal-to-noise ratio (SNR) of gradient-echo based sequences.

Balanced Steady-State Free Precession (bSSFP) sequences provide high SNR and image contrast in relatively short scan times[26]. However, the bSSFP signal is extremely sensitive to off-resonance effects[27], either due to magnetic field inhomogeneities or chemical shift. This sensitivity is severe for frequencies around 1/TR and may cause signal nulling in the images known as banding artifacts[28], which can be mitigated by acquiring multiple bSSFP images[29–33], each with a different linearly increasing phase increment. Measurement of the bSSFP frequency response via phase-cycling, i.e. phase-cycled bSSFP, revealed profile asymmetries that indicated the presence of multiple tissue components (e.g. different types of tissue) within voxels[34]. Such asymmetries were observed in gray matter, white matter, and muscle[34–36], were stronger at high magnetic fields[37], and can be avoided by reducing TR[38]. Profile asymmetries were used to obtain diffusion metrics[39], and recent work suggests an interplay between the contributions of microstructural, chemical shift and chemical exchange effects to the asymmetry of the bSSFP profile[40]. Although phase-cycled bSSFP was used for water-fat signal separation[41–45], its potential to quantify fat by analyzing profile asymmetries has not been investigated.

The aim of this work was to develop a novel quantitative method for measuring fat fraction using bSSFP Signal Profile Asymmetries for Robust multi-Compartment Quantification (SPARCQ). By using asymmetries in bSSFP profiles, this method has the potential to overcome the limitations of current techniques. The proposed approach was tested in simulations, fat fraction phantoms, and healthy volunteers, and was compared with reference standard PDFF approaches such as $^1$H-MRS and multiecho GRE techniques.

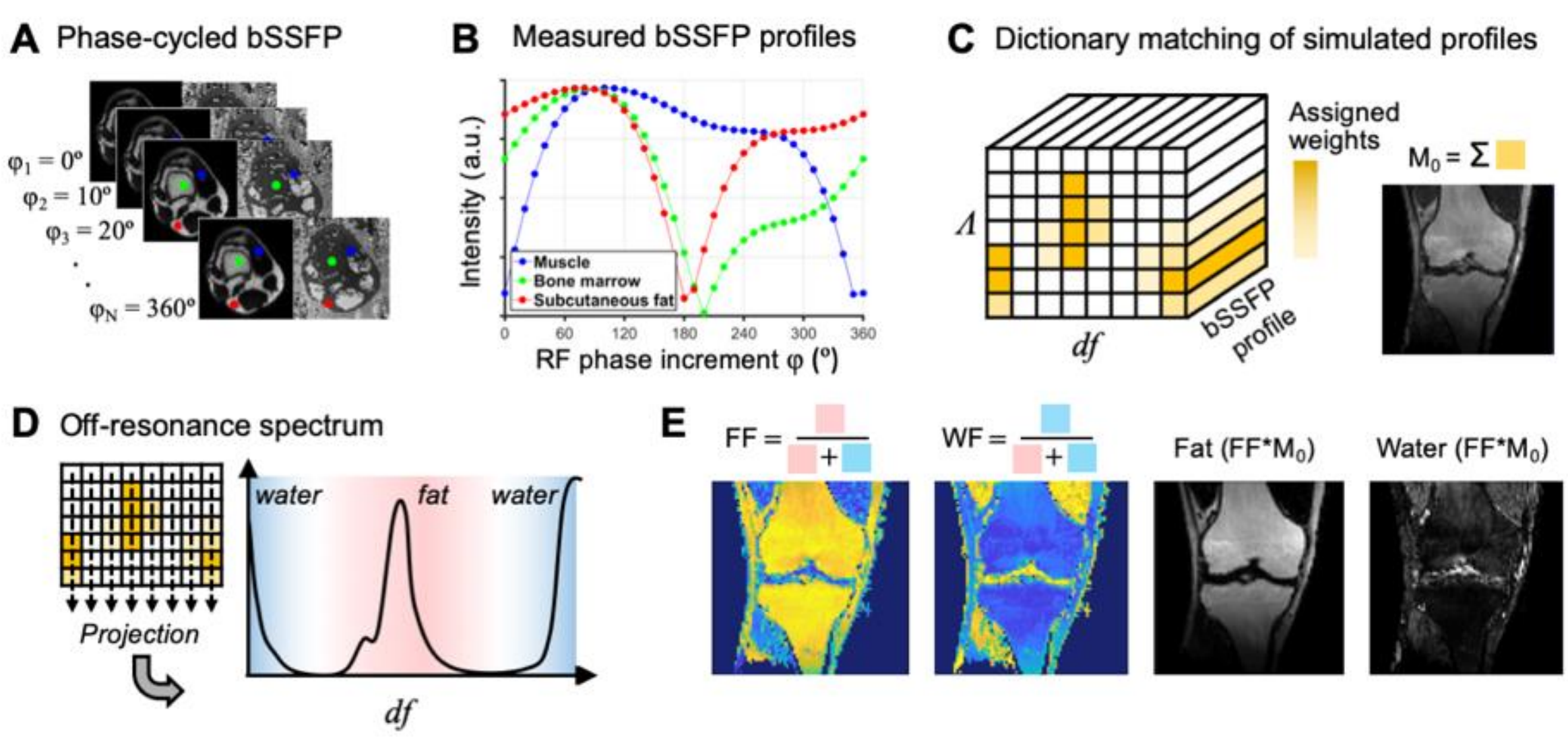

**Figure 1. Overview of the SPARCQ framework.** A phase-cycled bSSFP acquisition is performed (A) and for each voxel, a complex bSSFP profile is obtained (B). A dictionary of simulated bSSFP signal profiles is constructed, to which the measured bSSFP profile is matched (C), resulting in a weight matrix. Each weight represents the contribution of the corresponding simulated dictionary entry to the acquired profile. The equilibrium magnetization ($M_0$) is estimated by summing all the weights in the weight matrix. *df* spectra are obtained by projecting the weight matrix (D). The fat fraction (FF) is computed as the integral over the frequency range assigned to this component (panel D, red background) divided by the sum of the integrals of fat and water. Applying the same procedure to the whole imaging volume allows one to obtain quantitative maps. $M_0$-weighted images for fat and water are obtained by multiplying the fraction maps by the $M_0$ map.

## 2. METHODS

### *2.1 SPARCQ acquisition*

To measure phase-cycled bSSFP signal profile asymmetries, the SPARCQ acquisition (**Fig. 1A**) consists of multiple (N=36) bSSFP acquisitions, each with a different linear RF phase increment $\varphi_j$ according to:

$$\varphi_j = \frac{2\pi}{N}(j-1), \quad j=1, 2, \ldots, N \qquad \textbf{[1]}$$

Fully-sampled 3D Cartesian data were acquired using slab-selective excitation with the following parameters: TR/TE=3.4/1.7ms, RF excitation angle α=35°, a matrix size of 112x84, a field-of-view of $168\text{x}224\text{mm}^2$, with sufficient slices to cover the volume of interest, an isotropic resolution of $(2\text{ mm})^3$ and a receiver bandwidth of 930Hz/px. A TR of 3.4ms was deliberately chosen to maximize the difference between bSSFP profiles of water and fat (**Supporting Information Figure S1**), see also[46]. More intuitively, with a TR of 3.4ms, the locations of the water and main fat peak will wrap into the obtained off-resonance frequency spectrum (Fig. 1D) in such a way, that their relative distance is maximized. The choice of TR is fundamental in preventing signals from different components to overlap when wrapping in the [0, 1/TR] range.

At the beginning of each phase-cycled acquisition, ten ramp-up pulses were applied with linearly increasing RF power[47]. The total acquisition time was 20:35 minutes. Magnitude and phase images were reconstructed on the scanner and DICOM files were exported.

## *2.2 SPARCQ reconstruction*

The SPARCQ reconstruction can be divided into five steps that were all implemented in Matlab 8.5 (MathWorks, Natick MA):

1. Construction of a complex bSSFP profile for each voxel using measured data (**Fig. 1AB**).
2. Creation of a dictionary containing simulated phase-cycled bSSFP profiles. Each entry in the dictionary has a specific off-resonance frequency *df* and a relaxation time ratio $\Lambda$ ($T_1/T_2$) (**Fig. 1C**).
3. Matching of measured bSSFP profiles to a dictionary via a sum of simulated bSSFP profiles (**Fig. 1C**).
4. Computation of *df* distributions based on obtained weights (**Fig. 1D**).
5. Extraction of water and fat fraction maps and images (**Fig. 1E**).

*Step 1 – Measurement and extraction of complex phase-cycled bSSFP profiles*

Complex images were formed from the magnitude and phase DICOM images for each of the N scans. For each voxel, the complex bSSFP profile was then obtained by taking the complex values as function of applied phase increment (**Fig. 1A-B**).

*Step 2 – Creation of the dictionary*

The dictionary contains a simulated phase-cycled bSSFP profile for each combination of relaxation time ratio $\Lambda$ and off-resonance frequency *df*, resulting in a 2D dictionary (*Λ,df*) of bSSFP profiles. Profiles were simulated using the Bloch equations with RF phase increments $\varphi_j$, TR, TE and RF excitation angle α that are identical to the sequence parameters (**Section 2.1**). The resolution of the dictionary was manually tuned. When approaching the limit $TR/T_2 \rightarrow 0$, relaxation time properties enter the dictionary uniquely via their ratio Λ, which was varied from 1 to 26 in steps of 5. This results in a wide enough range of $T_1$ and $T_2$ values, so that the water and fat relaxation times are both represented. The off-resonance dimension *df* included any frequency within the detectable 1/TR range, from 0Hz to 1/TR~294Hz in 8Hz steps. This dimension represents both contributions from magnetic field inhomogeneities and chemical shift. Constraints on chemical shifts and peak amplitudes, such as those used in multiecho approaches, were not used. Instead, the distinction between water and fat is made based on their matched position in the off-resonance dimension.

*Step 3 – Dictionary matching*

The presence of different tissue fractions shapes the measured bSSFP profile, which is a complex sum of individual bSSFP profiles, providing a means of extracting fractional information. However, hardware-related factors can introduce an offset between the phase of the acquired bSSFP profile and the phase of the best dictionary fit, creating discrepancies between measured and simulated profiles, which may affect dictionary matching accuracy. To address this issue, a phase augmentation step is performed to guide the final matching process. Following the fingerprinting approach[48], the profile that minimizes the dot product between the dictionary entry and the acquired bSSFP profile is selected, and the median difference with the phase profile of the acquired bSSFP profile is used to augment the phase of the acquired profile. Phase augmentation is a rapid initial estimation which does not require high precision. Moreover, it has negligible influence on fat fraction estimation (**Supporting Information Methods S1 and Supporting Information Figure S2**).

Next, the complex bSSFP profile is expressed as a weighted sum of dictionary entries. This approach allows detecting multi-compartment compositions within each voxel and provides information on the contribution of each dictionary entry to the measured signal. This technique builds upon previous work on myelin-water imaging[49,50], but extends it to two dimensions, $\Lambda$ and $df$. To accomplish this, a two-dimensional weight matrix, the same size as the dictionary, is defined in which each cell represents the weighting of the corresponding dictionary entry (**Fig. 1C**).

All profiles are expressed as a concatenation of their real and imaginary parts, to guarantee real-valued weights, doubling the length of each signal to $2N$. To further simplify the fitting, the two-dimensional dictionary of bSSFP profiles and the two-dimensional weight matrix are reshaped to a $2N \times k$ matrix for the dictionary and a $k \times 1$ vector for the weights. Where $k$ has length of 216 (6 $\Lambda$ values x 36 $df$ values). By naming $\mathbf{s}_{acq}$ the $2N \times 1$ acquired signal, $\mathbf{D}$ the $2N \times k$ reshaped dictionary and $\mathbf{w}$ the $k \times 1$ reshaped weight matrix, the optimization problem can be mathematically described as follows:

$$\hat{\mathbf{w}} = \underset{\mathbf{w}}{\operatorname{argmin}} \ \{\left\|\mathbf{D} \cdot \mathbf{w} - s_{acq}\right\|_2^2 + \lambda\|\boldsymbol{\Delta}\mathbf{w}\|_2^2\} \text{ subject to } \mathbf{w} \geq 0 \qquad \textbf{[2]}$$

The $L_2$ norm was chosen as a distance metric and the constraint $\mathbf{w} \geq 0$ was added to avoid negative weights that would hinder finding a biologically plausible solution. Therefore, the first term corresponds to a classical non-negative least-squares (NNLS) problem. Since the problem is ill-posed and prone to overfitting, a second order Laplacian regularization term was added to favor smoothness between neighboring weights $\boldsymbol{\Delta}\mathbf{w}$, as described previously[49,50]. The solution of the NNLS problem is the *k* x *1* array of weights, which can be used to compute the *2N* x *1* best fit signal given by the weighted sum of all the signals in the dictionary (**D·w**). The final weight matrix is obtained by representing the array of weights in its original 2D shape, with relaxation time ratio and off-resonance frequencies as dimensions. The regularization parameter was manually tuned and tested (**Supporting Information Figure S3**) and set to 0.08 in phantom experiments and to 0.025 in volunteer experiments.

Consequently, the sum of the weights in a voxel does not necessarily add up to 1, but to the total signal intensity in that voxel. The weights only represent the relative contribution of each compartment to the total signal. It is important to note that the signals are not normalized in the multi-compartment fitting step, as this can introduce errors in the estimation of the weights.

*Step 4 - Estimation of off-resonance spectra*

The weight matrix contains information on the voxel content in terms of relaxation time ratios and off-resonance frequencies. To visualize the off-resonance information, distributions of the off-resonance frequency content are obtained by projecting the weight matrix onto the axis of off-resonance frequencies (**Fig. 1D**).

*Step 5 – Quantitative parameter mapping and qualitative image reconstruction*

The composition of a given voxel can be extracted from the obtained spectrum (i.e., distribution of weights). For a 2-compartment system, the fraction of one tissue component with a specific frequency range (e.g., fat) can be computed as the integral of the projected *df* spectrum over the frequency range assigned to it, divided by the sum of the integrals over the frequency ranges assigned to the first and second component (**Fig. 1E**).

In the case of water-fat separation, the fat fraction (FF) and water fraction (WF) are defined as:

$$\mathrm{FF} = \frac{\mathrm{F}}{\mathrm{F+W}} \quad [3]$$

$$\mathrm{WF} = 1 - \mathrm{FF} \quad [4]$$

where F and W are the integrals of the *df* spectrum over the frequency ranges assigned to fat and water, respectively.

To account for small frequency shifts caused by local magnetic field inhomogeneity, the peak that is closest to either 0Hz or 1/TR is assigned to water. The spectrum is then shifted along the off-resonance axis so that the assigned water peak is centered around 0Hz. Next, the position of the fat peak is determined, which is defined as the peak with the largest amplitude between 0 and 1/TR. The first minima between the water and the fat peak are used as integration limits to obtain F and W.

In a voxel, the thermal equilibrium magnetization $M_0$ can be estimated as a weighted sum of each dictionary entry since the weight matrix gives an estimation of the contribution of each signal in the dictionary to the acquired signal. As the signals in the dictionary were all simulated for $M_0$=1, the thermal equilibrium magnetization in a voxel can be estimated as the sum of all weights in the matrix.

Application of the framework over the whole volume (N.B.: step 2 of dictionary creation only needs to be completed once for the whole volume), quantitative maps of FF, WF, and equilibrium magnetization $M_0$ can be obtained (**Fig. 1F**).

The framework allows the extraction of qualitative images of individual components that may be easier to interpret (**Fig. 1F**). To do this, $M_0$-weighted images for each component can be obtained by voxel-wise multiplication of the corresponding fraction maps, in the current case WF or FF, with the $M_0$ map.

### *2.3 Numerical simulations*

Numerical simulations were performed to evaluate the accuracy of SPARCQ for fat fraction quantification and to test its robustness in the presence of noise.

First, $^1$H NMR spectra were simulated that consisted of a single water peak resonating at 0Hz and two fat peaks at -485Hz and -434Hz, with a relative 1:4 peak area ratio. Each peak was modeled as a Lorentzian function with a full width half maximum (FWHM) of 20Hz according to

$$L(x) = \frac{A}{1+x^2} \quad \text{with } x = \frac{2(P-P_0)}{FWHM} \qquad [5]$$

with A the peak area, P the frequency ranging from -600Hz to 200Hz with steps of 1Hz, and $P_0$ the center frequency. The three peak areas were adjusted for different FF ranging from 0 to 1 in steps of 0.05. Then, the three Lorentzian line shapes were summed to obtain a set of simulated $^1$H spectra for each FF.

Second, Bloch simulations were performed to generate bSSFP signal profiles for each frequency of the simulated $^1$H spectra. Off-resonance frequencies were varied from -600Hz to 200Hz (in steps of 1Hz), with all other parameters being fixed (TR=3.40ms, TE=1.7ms, $\alpha$=35°, $\Lambda$= 6 [$T_1$=480ms, $T_2$ =80ms]). To obtain a final complex bSSFP profile for each FF, the simulated bSSFP profiles were weighted by the amplitude of their corresponding off-resonance frequency in the simulated $^1$H spectrum and summed.

Third, randomly generated white Gaussian noise (SNR levels ranging from 5 to 100 in steps of 2.5) was added to the imaginary and real part of each bSSFP profile. SPARCQ was used for FF estimation in each of the generated bSSFP profiles. The experiment was iterated 100 times, using a different seed at each iteration for the randomly generated noise. The mean and standard deviation of the estimation error over the 100 iterations were calculated. This experiment was repeated to confirm the negligible effect of the phase augmentation step on FF estimates (**Supporting Information Methods S1**).

### *2.4 Fat fraction phantom experiments at 3T and 9.4T*

The accuracy of the proposed framework was evaluated in phantoms. First, a dedicated fat-water phantom was created[51] composed of six 50mL Falcon tubes with different partial volumes of peanut oil (0%, 20%, 40%, 60%, 80%, 100%) immersed in a 3% weight agar solution (**Figure 4C**). Peanut oil was chosen because it has resonance frequencies similar to those of triglycerides in human adipose tissue[25]. To quantify the PDFF, unlocalized $^{1}H$ MRS was performed separately in each Falcon tube in a 9.4T MRI scanner (Magnex Scientific, Oxford, UK) with a Direct Drive spectrometer (Agilent, Palo Alto, CA, USA). Line shape fitting and integration in jMRUI4.0 was used to calculate the peak areas of all visible resonances and obtain the gold standard FF(GS). In addition, a chemical shift thresholded (CST) fat fraction was quantified for each tube. The CST method assigns resonance peaks between 3.5ppm and 6.0ppm to water and resonance peaks between 0ppm and 3ppm to fat. It was used as an additional metric for comparison because the current SPARCQ implementation assigns resonance peaks of fat close to the water resonance to the water compartment. This additional metric evaluates SPARCQ for tissue fraction quantification prior to using potential correction methods such as a 6-peak fat model[25].

Phantom acquisitions were performed on a clinical 3T scanner (MAGNETOM Prisma$^{fit}$, Siemens Healthcare, Erlangen, Germany) using a commercially available 18-channel body coil. A prototype 3D Cartesian phase-cycled bSSFP sequence (**section 2.1**) and a reference multiecho GRE (ME-GRE) sequence were used for data acquisition. ME-GRE acquisition parameters were: 13 monopolar echoes with echo times from 1.34ms to 25.10ms and an echo spacing of 1.98ms, an isotropic resolution of 2.0x2.0x2.0(mm)$^3$, RF excitation angle of 20°, a receiver bandwidth of 970Hz/pixel, and an acquisition time of 3:02min. ME-GRE data were processed with the ISMRM Fat-Water Toolbox[15,16] to obtain fat fractions, using a 9-peak fat spectral model determined by the 9.4T MRS experiments. $B_0$ field maps provided by the Toolbox were used to estimate magnetic field inhomogeneity.

Fat fractions were quantified in manually drawn regions of interest (ROI) corresponding to the six vials. FF obtained with SPARCQ and ME-GRE were compared with the gold standard FF determined with $^{1}H$ MRS at 9.4T. ROIs were chosen to span five consecutive slices for comparability with unlocalized spectroscopy data. To focus on accuracy of SPARCQ and not on its sensitivity to $B_0$ inhomogeneities, the five slices per vial were chosen in regions where $B_0$ inhomogeneities in the agar surrounding the vials was as small as possible. Additional analyses were performed to determine the impact of $B_0$ inhomogeneities on SPARCQ-based FF estimates, and to identify sources of error that can be mitigated in future implementations. To test this, $B_0$ information extracted from the ME-GRE scans was used to shift the acquired phase-cycled bSSFP profiles prior to dictionary matching (**Supporting Information Methods S2**).

### *2.5 Volunteer experiments*

Volunteer experiments were performed in knees using a commercially available 15-channel Tx/Rx knee coil (Quality Electrodynamics, Mayfield, OH, USA) at the same clinical 3T scanner as described before. The volunteers gave written and informed consent. The study was carried out according to institutional rules.

*Scan-rescan repeatability study in volunteers*

A scan-rescan repeatability study was performed on six human knees, with repositioning in between scans. For reference, a Dixon acquisition was performed with a turbo spin echo (TSE) Dixon sequence with TR/TE=3470/100ms, RF excitation angle $\alpha$=150°, matrix size 128x80, field of view 160x256mm, isotropic resolution $(2mm)^3$, receiver bandwidth 601Hz/px and turbo factor 18. In-phase, out-of-phase, water and fat images were reconstructed on the scanner.

Quantitative maps were obtained with SPARCQ as described. A thresholding was performed based the amplitude range of the profile, which was performed to mask the background and voxels with low signal.

Spectra and estimated parameters were checked in areas close to air-tissue interfaces (e.g., patella) to evaluate the efficacy of SPARCQ in regions of strong $B_0$ inhomogeneity. Bland-Altman analysis was performed on the scan-rescan data on the mean fat fractions obtained with SPARCQ in elliptical ROI in five different tissues (vastus medialis, biceps femoralis, semimembranous muscles, bone marrow, and subcutaneous fat). Water-fat-separated images obtained with SPARCQ were compared with those obtained with Dixon scans.

*Comparison with ME-GRE*

A comparison was made between the FF estimates obtained with SPARCQ and ME-GRE on two human knees. Phase-cycled bSSFP scans as well as ME-GRE scans were performed as described earlier. For data processing with the ISMRM Fat-Water Toolbox, a 6-peak fat spectral model was used[25]. Quantitative maps were obtained with SPARCQ and ME-GRE, and a Bland-Altman analysis was performed on the mean FF obtained with SPARCQ and ME-GRE in elliptical ROIs in five different tissues (vastus medialis, biceps femoralis, semimembranous muscles, bone marrow, and subcutaneous fat).

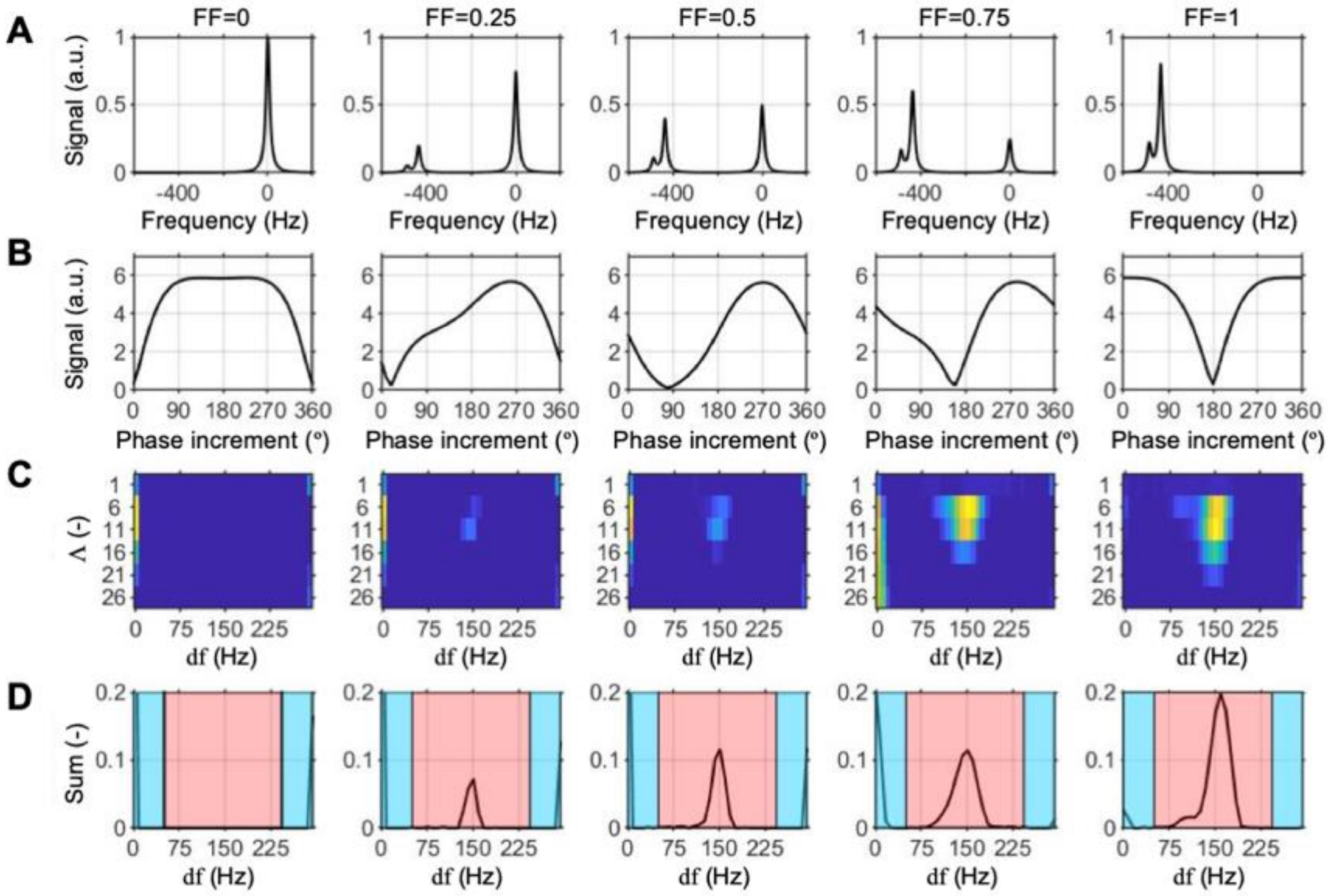

**Figure 2. Simulation experiment.** [1]H MR spectra were simulated in the frequency range -200 to 600 Hz for fat fractions going from 0 to 1 in steps of 0.25 (A). The corresponding phase-cycled bSSFP signal profiles were simulated (B) (magnitude). Following dictionary matching, weight matrices are obtained (C), which are used to obtain off-resonance frequency spectra with a bandwidth ranging from 0 to 1/TR (D). The locations of water and fat peaks (C,D) are wrapped within a 1/TR frequency range.

# 3. RESULTS

## *3.1 Numerical simulations*

bSSFP signal profiles (**Fig. 2B**, only magnitude is shown) obtained from simulated [1]H MR spectra (**Fig. 2A**) show a clear difference in shape when the fat fraction is changed from 0 to 1, whereas signals from mixed water and fat display large asymmetries. The off-resonance frequency spectra (**Fig. 2D**) derived from the weight matrix (**Fig. 2C**) correlate well with the simulated spectra. It should be noted that whilst the simulated spectra cover a large frequency range, the spectra derived from SPARCQ are limited to the range 0-1/TR. Therefore, a wrapping of the spectra can be observed. The fat resonance frequencies around -440Hz in the simulated spectra are found at around 150Hz in the off-resonance frequency spectra derived from the fitting. This is reasonable, since unwrapping the phase advance accumulated in a TR of 3.4ms for a frequency of -440Hz leads to a phase advance of 178°, corresponding to a frequency of 148Hz.

The fat fractions estimated with SPARCQ (**Fig. 3A**) in the presence of different noise allowed to evaluate the accuracy and noise robustness of the proposed algorithm. The mean fitting error and its standard deviation over the 100 repetitions (**Fig. 3B** and **Fig. 3C**) showed that strong noise (SNR<20) leads to reduced accuracy (**Fig. 3B**) and precision (**Fig. 3C**), which is confirmed by both the higher mean and the higher standard deviation of the error over repetitions. In noisy signals (SNR<10), small fat fractions (~0.2) seem to be more accurately estimated than higher ones (~0.8), the latter demonstrating severe underestimations. For higher SNR (>20) estimations gain in both accuracy and precision. However, small underestimations (for FF>0.7 and FF<0.15) or overestimations (for 0.15<FF<0.7) were observed.

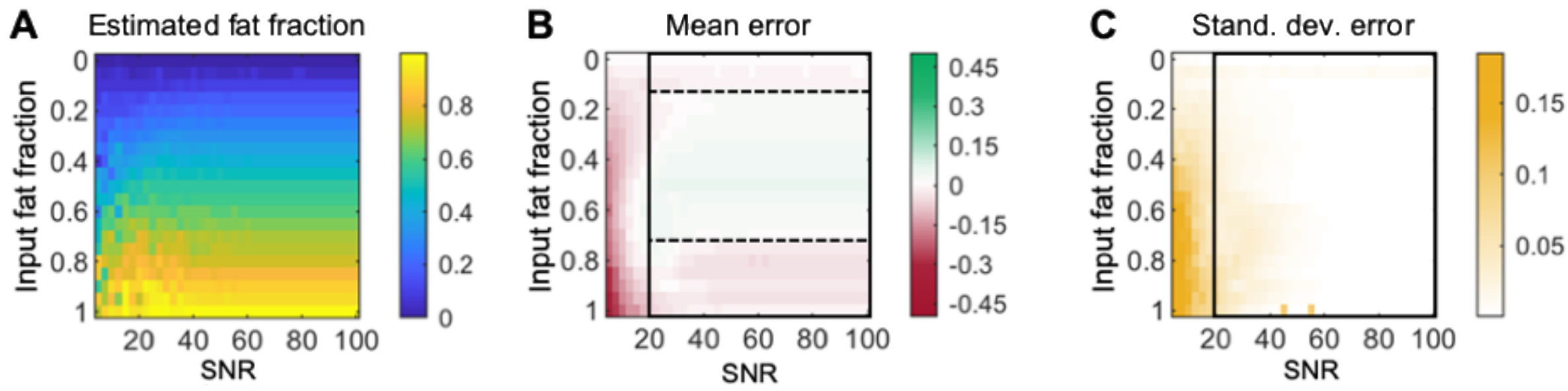

**Figure 3. Simulation experiments to test the noise robustness of SPARCQ.** Fat fractions estimated with SPARCQ depending on the SNR (A). Mean error of the fat fraction estimation (estimated-input) depending on the SNR for 100 repetitions (B). Standard deviation of the fat fraction estimation error depending on the SNR for 100 repetitions (C).

### *3.2 Phantom experiments at 3T and 9.4T*

$^{1}$H spectra acquired at 9.4T in the different Falcon tubes showed well-resolved resonance frequencies (**Fig. 4A**). Besides one tube, which was excluded, all phantom components formed neat agar emulsions without phase separation of water and oil. The gold standard fat fractions FF(GS), determined with $^{1}$H MRS, were 0%, 22%, 42%, 64%, 100%. The corresponding fractions assigned using a chemical shift threshold, FF (CST), were 0%, 20%, 39%, 58%, and 91% respectively.

Fat fraction quantification with ME-GRE agreed well, with a regression line FF(ME-GRE)=0.947*FF(GS)+0.074 (**Fig. 4B**). SPARCQ fat fraction quantification (**Fig. 4C**) showed good agreement with a regression line FF(SPARCQ)=0.861*FF(GS)+0.046 (Fig. 4E), and a regression line FF(SPARCQ)=0.947*FF(CST)+0.044 (Fig. 4D). The visible bias is mainly driven by vial V1, which had a strong $B_0$ field inhomogeneity (~60Hz) (**Fig. 4B**, color bar). A $B_0$-weighted regression (**Fig. 4D**, blue regression line) allowed to correlate the lower precision and higher standard deviation of fat fractions estimated with SPARCQ in vial V1 and V4 to $B_0$ field inhomogeneities.

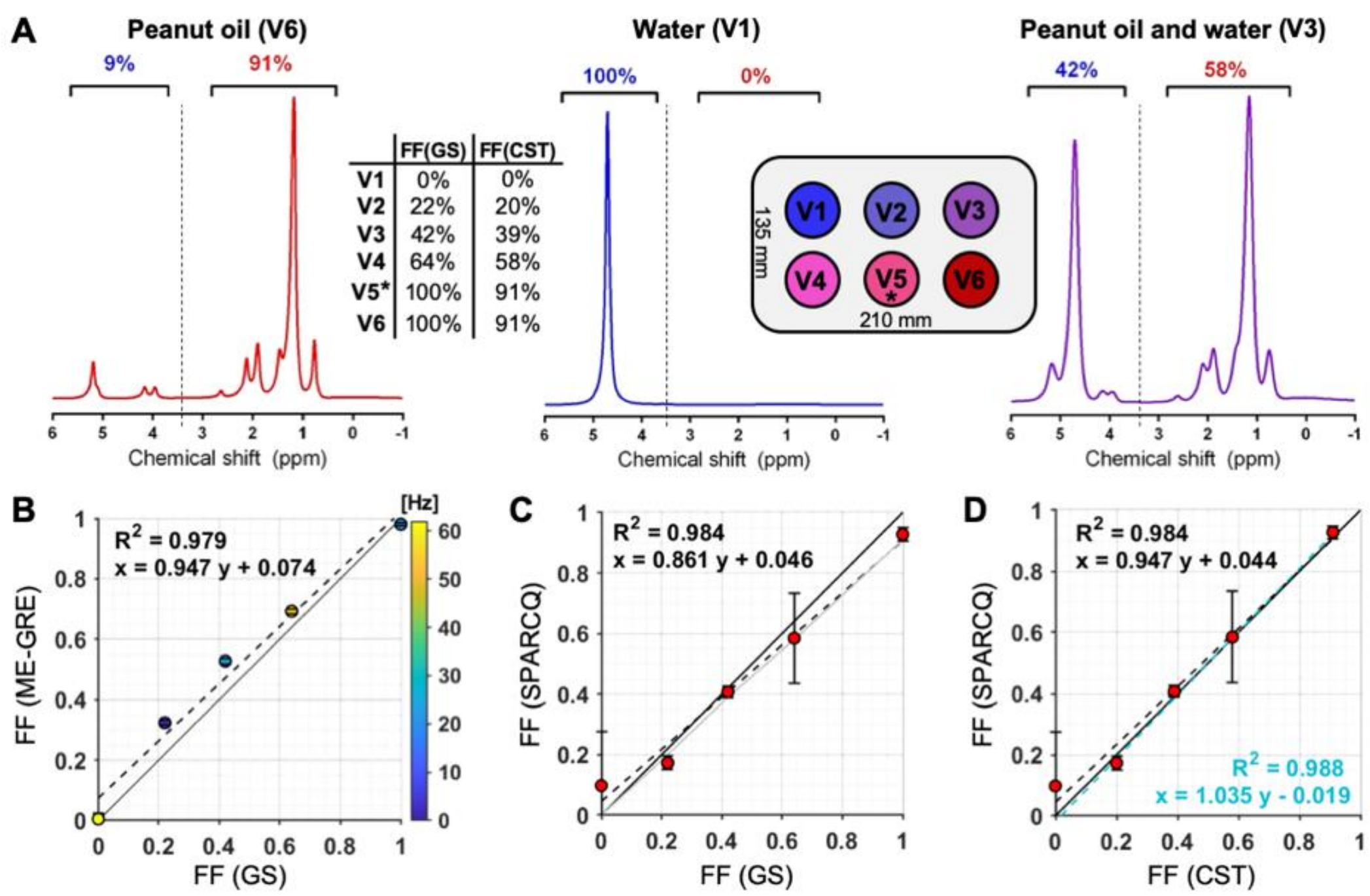

| | FF(GS) | FF(CST) |
|---|---|---|
| V1 | 0% | 0% |
| V2 | 22% | 20% |
| V3 | 42% | 39% |
| V4 | 64% | 58% |
| V5* | 100% | 91% |
| V6 | 100% | 91% |

**Figure 4. Accuracy of SPARCQ. (A)** Example of $^1$H MR spectra obtained at 9.4T for three vials in the fat phantom. Peak areas were obtained for all visible resonances using jMRUI, which were then used to calculate the gold standard proton density fat fraction FF(GS) and a chemical shift threshold fat fraction FF (CST). For FF (CST), resonances above 3.5ppm were assigned to water, below 3.5ppm to fat (dashed vertical line in **panel A**). Fat fractions estimated with ME-GRE (**B**) and SPARCQ (**C,D**) in each vial versus FF(GS) and FF(CST). The unit line (solid black line) and the regression line (dashed black line) are also plotted. In **panel B**, the color code indicates the mean $B_0$ field inhomogeneity (in Hz) within the ROI considered. Field inhomogeneity is the cause for the SPARCQ fat fraction deviation (vial V1) and increase in standard deviation (vial V4). In **D**, the light blue regression line was obtained by weighting each sample by the inverse of the mean $B_0$ field inhomogeneity. *V5 was excluded because of water-oil phase separation shortly after phantom building.

### *3.3 Volunteer experiments*

All quantitative maps (**Fig. 5A, Fig. 6**) and water-fat-separated images (**Fig. 5B**) obtained with SPARCQ demonstrated the expected contrast between fatty (bone marrow, subcutaneous fat) and non-fatty (muscles) tissues. In peripheral regions close to air-tissue interface (e.g., patella), an inverted contrast may be observed (**Supporting Information Figure S4, S5**). Despite the difference in contrast in water-fat-separated images obtained with SPARCQ ($M_0$ weighting) and with Dixon ($T_2$ weighting) (**Fig. 5B**), SPARCQ fat and Dixon fat images were in good agreement based on a visual assessment. In water images, hyperintensities were observed in the same structures filled with liquid.

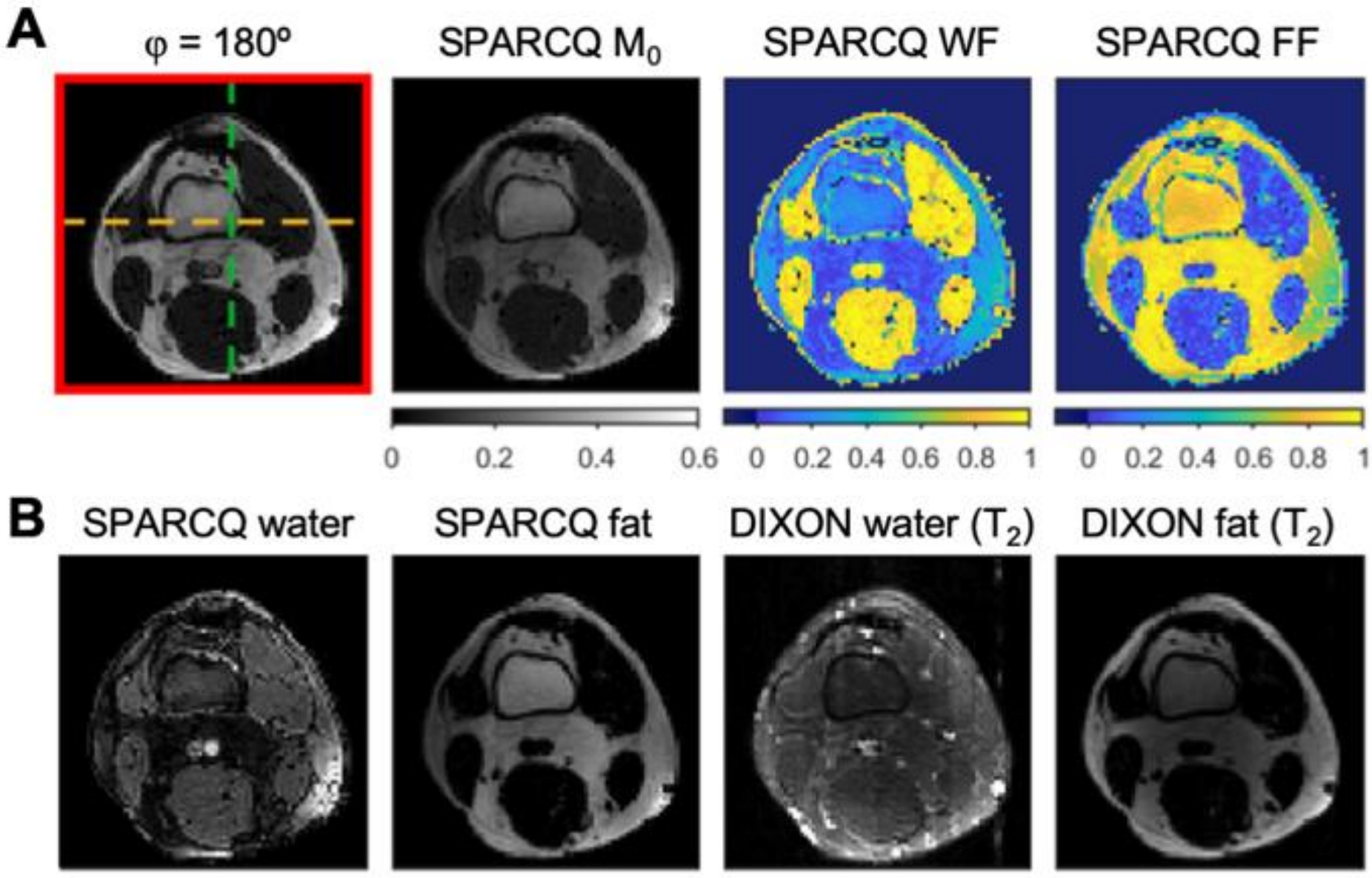

**Figure 5. Quantitative maps and water-fat-separated images.** Example images of one volunteer. (A) Quantitative maps obtained with SPARCQ in an axial view. (B) $M_0$-weighted water-fat-separated images obtained with SPARCQ and T2-weighted water-fat-separated images obtained with TSE Dixon (TR/TE=3470/100 ms, resolution $(2mm)^3$, receiver bandwidth 601Hz/px, turbo factor 18; in-phase, out-of-phase, water and fat images were reconstructed on the scanner). **Supporting Information Figure S4** shows the axial, sagittal, and coronal views of this data set.

Quantitative maps obtained with SPARCQ and ME-GRE (**Fig. 6**) were in agreement, with ME-GRE showing more polarized (i.e., 100% fat or 100% water) estimations. In peripheral regions close to the air-tissue interfaces (e.g., patella), ME-GRE outperformed SPARCQ. Bland-Altmann analysis on the fat fraction maps estimated with SPARCQ and ME-GRE (**Fig. 6**) revealed limits of agreement (LOA) of -0.0170±1.96*0.0665 between the two methods, with greater disagreement in voxels where SPARCQ was estimating more mixed tissue compositions (e.g., bone marrow).

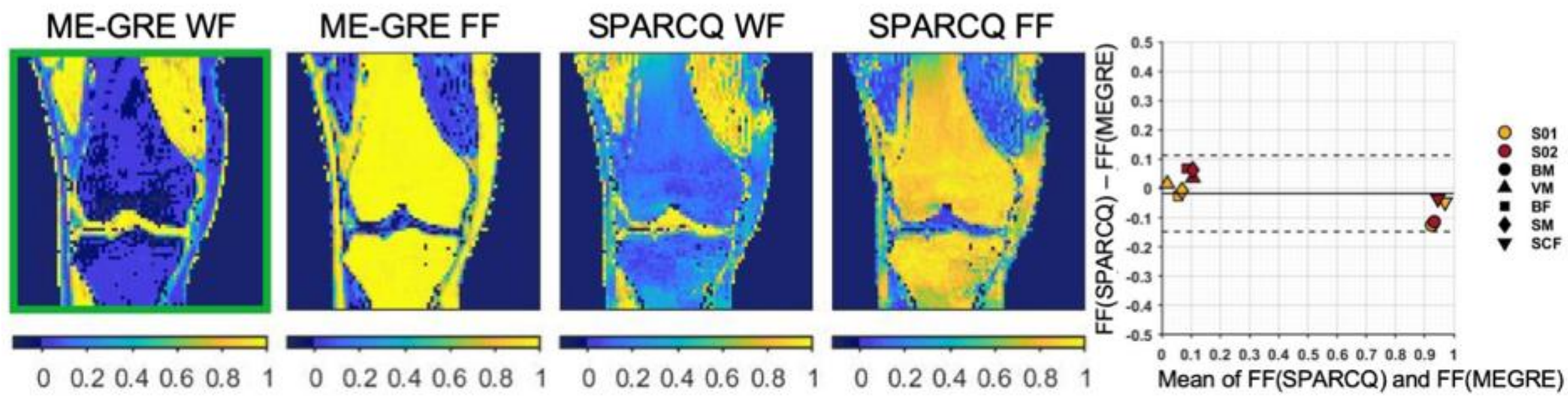

**Figure 6**. Quantitative water and fat fraction maps obtained with phase-cycled bSSFP (SPARCQ) and multiecho GRE (ME-GRE) in one volunteer. Bland-Altmann analysis of fat fractions estimated with SPARCQ and ME-GRE in 2 volunteers (1 volunteer=1 color) for 5 regions of interest (1 marker

type =1 region). Limits of agreement: LOA = -0.0170 ± 1.96*0.0665. BM : *bone marrow*, VM : *vastus medialis*, BF : *biceps femoralis*, SM : *semimembranous muscle*, SCF : *subcutaneous fat*. **Supporting Information Figure S5** shows the axial, sagittal and coronal views of this data set. The current SPARCQ implementation does not consider $B_0$ inhomogeneities, resulting in variations close to air interfaces.

Off-resonance frequency spectra obtained from the weight matrices in fatty regions (**Fig. 7A**) and non-fatty regions (**Fig. 7B**) far from the air-tissue interface showed the expected frequency components: a narrow on-resonant peak corresponding to water, and a broader off-resonant peak corresponding to fat. On the other hand, in a region near an air-tissue interface (as shown in **Fig. 7C**), water and fat were misclassified due to local $B_0$ inhomogeneity and the rigid assignment of the water peak in the current implementation.

Fat fractions in ROIs in different tissues (**Fig. 7B**) showed good agreement between all the subjects, with muscles showing a low fat content (3.2±0.9% vastus medialis, 7.3±2.8% biceps femoralis, 5.8±1.5% semimembranous muscles) and bone marrow and subcutaneous fat confirming their high fat content (81.7±2.4% bone marrow, 94.3±0.9% subcutaneous fat).

The Bland-Altmann analysis on the fat fraction maps in the scan-rescan experiment (**Fig. 8**) revealed very low bias b=0.0074 and a good coefficient of repeatability CR=0.0512.

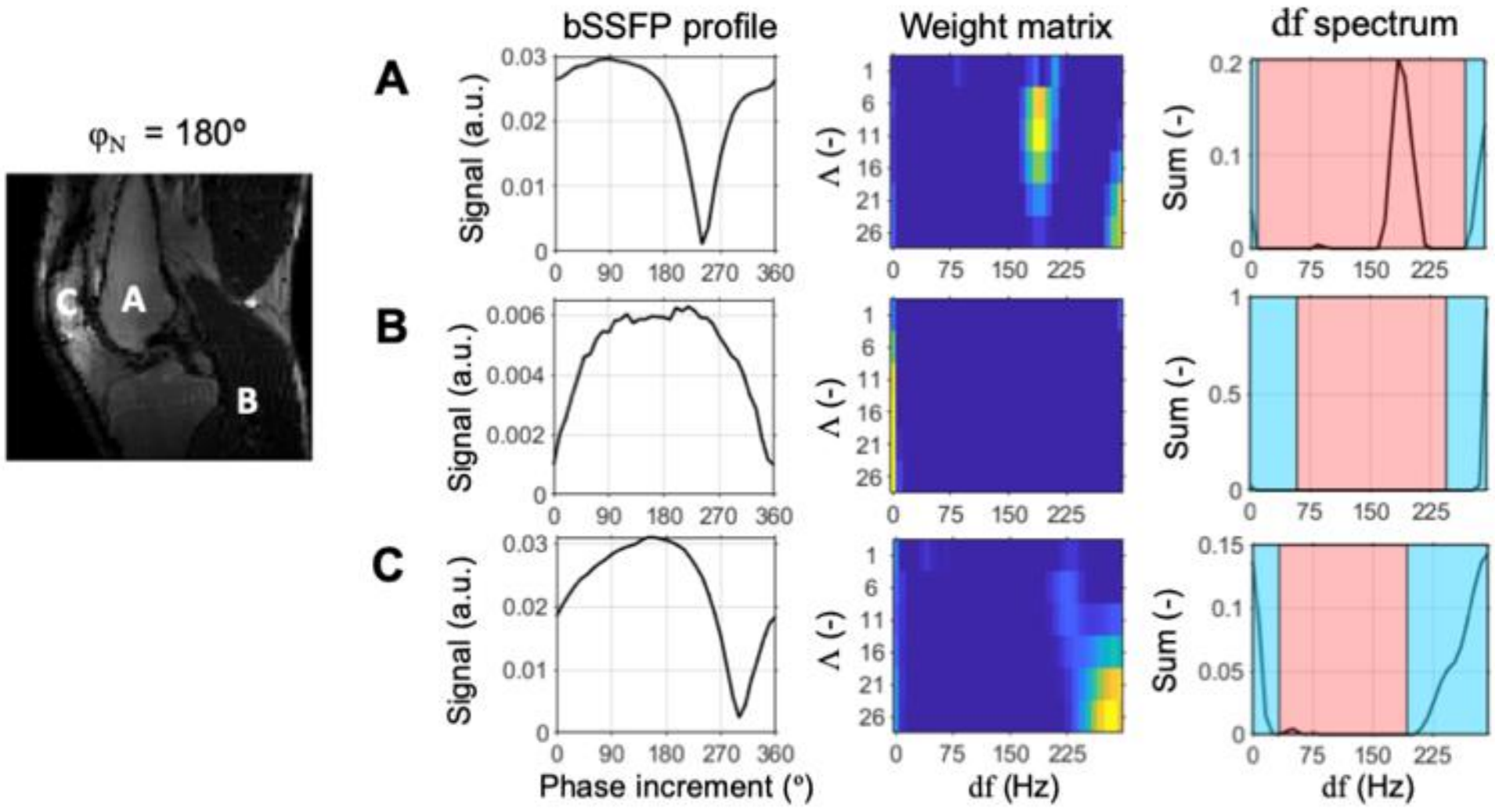

**Figure 7. Dictionary matching results in three voxels belonging to different tissues**. Acquired bSSFP signal profiles (**left column**), optimized weight matrices (**middle column**) and off-resonance frequency spectra (**right column**) for a voxel belonging to fatty (bone marrow, **A**), non-fatty (muscle, **B**) tissue and a voxel close to the air-tissue interface (patella, **C**). The *df* spectrum (**right**) is obtained by projecting the weight matrix onto the $df$ axis (SOW: sum of weights). The selected ranges of tissue-

specific frequencies are marked blue (fat) and red (water). In bone marrow (**A**), multiple frequency components are detected (on-resonant water, off-resonant fat). In muscle (**B**), a single frequency component (on-resonant water) is detected. Close to the air-tissue interface (**C**), the peak in the *df* spectrum is wrongly assigned to water presumably due to large $B_0$ field inhomogeneity. **Supporting Information Figure S6** demonstrates the capability of SPARCQ when such $B_0$ effects can be mitigated.

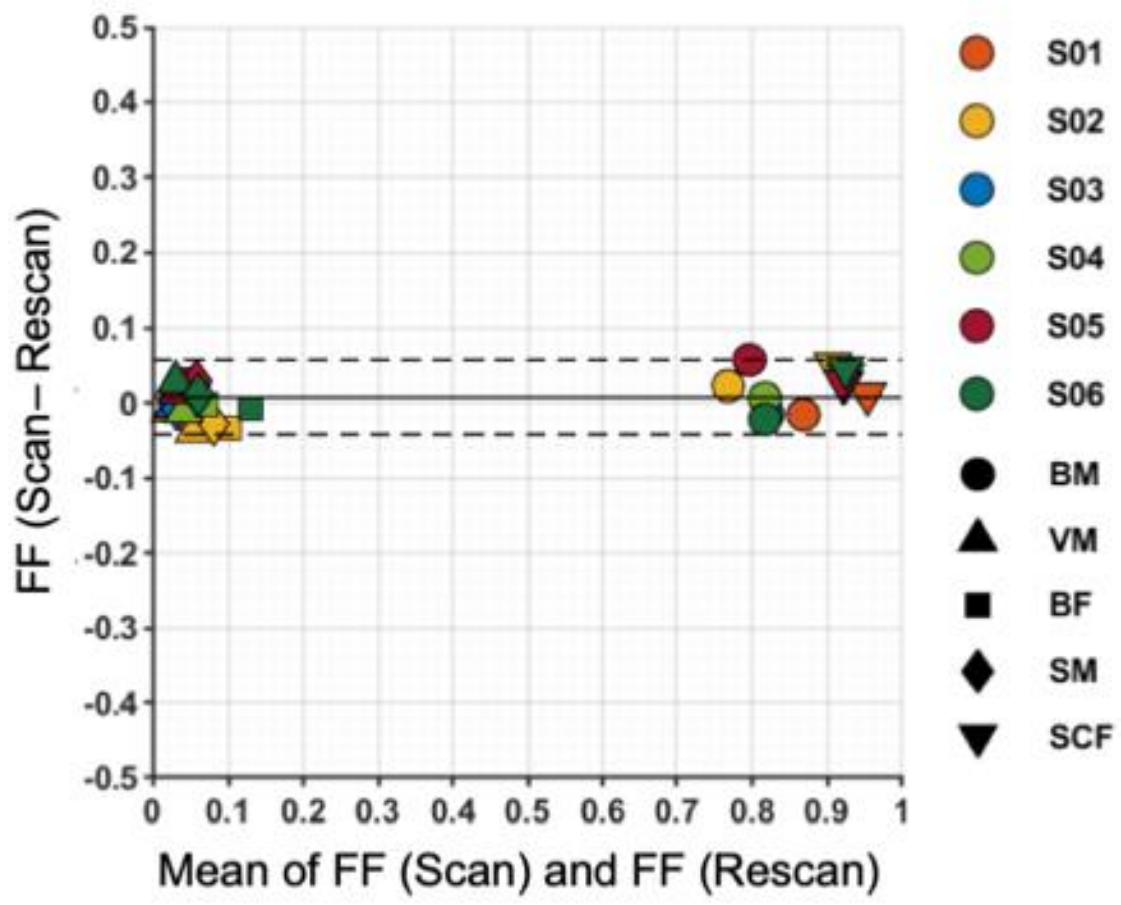

**Figure 8. Scan-rescan repeatability of the fat fractions estimated with SPARCQ.** Bland-Altmann analysis of scan-rescan repeatability in 6 volunteers (1 volunteer = 1 color) for 5 regions of interest (1 marker type = 1 region). Limits of agreement: LOA = 0.0074 ± 1.96*0.0574. Coefficient of repeatability: CR = 0.0512. BM: *bone marrow*, VM: *vastus medialis*, BF: *biceps femoralis*, SM: *semimembranous muscle*, SCF: *subcutaneous fat*.

## 4. DISCUSSION

This work presents a novel quantitative method for detecting different tissue compartments based on phase-cycled bSSFP profile asymmetries. SPARCQ estimates off-resonance frequency spectra from acquired bSSFP profiles in each voxel. The obtained spectra reveal the presence of different tissue compartments, such as water and fat, and allow the quantification of their individual contribution to the overall signal profile within the same voxel. In addition, SPARCQ provides an estimation of the equilibrium magnetization $M_0$, enabling the generation of distinct $M_0$-weighted images for each component, free of banding artifacts.

The proposed SPARCQ framework was demonstrated and validated for fat fraction mapping using comparisons with reference standards for PDFF, including high resolution $^1$H MRS and multiecho GRE. In phantom studies, both the ME-GRE and SPARCQ methods provided fat fraction estimates that were equally close to the ground truth values. However, the water-fat separation algorithm used with the ME-GRE data relied on a fixed fat spectrum measured in the phantom at 9.4T, whereas SPARCQ did not use a model with fixed peak locations and amplitudes. In the current implementation, SPARCQ assigns the three fat resonances around the water peak to the water compartment, resulting in an underestimation of fat that increases linearly with increasing fat fraction. Since fat peaks that resonate close to water are a fixed proportion of the total peak area of fat, straightforward corrections could be implemented in future work. The possibility of such a correction was demonstrated by the agreement of SPARCQ fraction estimates with a fraction based on a chemical shift threshold, the FF (CST) (Fig. 4C versus Fig. 4D). The CST metric was introduced to reflect better the performance of SPARCQ in the absence of a spectral fat model. Furthermore, it demonstrated that SPARCQ is capable of distinguishing compartments based on their underlying resonance frequency, and thus has the potential to accurately quantify fat fraction.

Volunteer experiments confirmed that SPARCQ, as expected in the current implementation, underestimated fat fraction compared with ME-GRE. These experiments also demonstrated the repeatability of SPARCQ. *In vivo* quantitative maps showed the expected contrast between fatty and non-fatty regions. $M_0$-weighted water-fat-separated images obtained with SPARCQ compared well with those obtained with reference Dixon techniques. Nonetheless, SPARCQ maps and images showed that in some tissue regions close to air-tissue interfaces (e.g., in the patella) fat was misclassified. This was caused by the rigid peak selection strategy applied to the frequency spectra, which assumes that the peak closest to on-resonance is water, which fails in the presence of strong $B_0$ inhomogeneities. Future studies will focus on using additional SPARCQ characteristics, such as the peak width in the frequency spectrum, to facilitate a $B_0$ independent peak assignment of fat and water. For example, fat can typically be identified as a broader peak in the frequency spectrum within each voxel (**Fig. 7**). Alternatively, approaches similar to those used to mitigate strong $B_0$ inhomogeneities for water-fat separation in ME-GRE data could potentially be extended to SPARCQ[16].

The potential to quantify fat fractions using asymmetries in the bSSFP profile has several advantages over current techniques. The first is that bSSFP MRI provides a higher SNR compared with GRE-based MRI techniques. Secondly, because of the bSSFP $T_1/T_2$ -contrast, the potential $T_1$ signal bias is reduced. Thirdly, SPARCQ does not suffer from T2* signal decay, as observed in ME-GRE, and assumptions on T2* signal decay are not needed compared with current water-fat separation algorithms. The last advantage is that SPARCQ has the potential to quantify fat without using strict assumptions on the underlying fat spectrum, in terms of chemical shift of the expected fat peaks and their amplitudes.

Approximating an acquired signal profile as a weighted sum of multiple dictionary entries rather than finding the best matched single dictionary entry has recently been proposed[49,50], and was extended to two dimensions in the current work. This approach provides various benefits: 1) It minimizes the dimensionality of the dictionary, thus speeding up the processing and reducing the memory footprint of the algorithm. 2) It relies on simple signal models, where bSSFP profiles are simulated for a single combination of frequency (*df*) and relaxation time ratio ($\Lambda$), without requiring prior knowledge of the location, amount, and amplitudes of the different spectral components of water and fat. Nevertheless, the current dictionary matching approach is relatively slow in terms of reconstruction speed, and the current voxel-wise approach does not consider nor leverage neighboring voxel information. In addition, the phase augmentation step, albeit fast, could be incorporated in the matching process with a spatial smoothness constraint in future implementations. This, together with exploiting joint sparsity constraints[52] may help to accelerate the reconstruction and increase the resolution of the dictionary.

The current work was intended to demonstrate the feasibility of our concept rather than optimizing for speed or resolution. The bSSFP profile was measured by performing 36 fully-sampled phase-cycled bSSFP acquisitions. Therefore, the resulting acquisition time of 20 minutes is significantly higher compared to the ME-GRE acquisition of 3 minutes. Studies, similar as those performed in single compartments for quantitative mapping with PC-bSSFP[53], will need to be performed in order to define the minimum required phase-cycled scans for water-fat-quantification. Furthermore, the effect of changing TR was not investigated. To quantify fat through profile asymmetries, the TR should be well above 2.2ms and well below 4.5ms to preserve asymmetries in the bSSFP profile at 3T (**Supporting Information Figure S1**). Shortening scan time by TR reduction is thus feasible, but will affect the bandwidth of the resulting off-resonance spectrum. This illustrates an intricate relationship between asymmetry sampling and detection that requires further studies that fell outside of the scope of the present work. For example, asymmetry sampling could be improved by including signal profiles like those obtained with fast interrupted steady state sequences[54,55]. Asymmetry detection could be improved by optimizing the RF excitation angle to find a better trade-off between SNR of different tissue types, while preserving signal asymmetries. For example, the current study focused on fat estimation and displayed rather low SNR in the water-only images.

The current study demonstrates the potential to quantify fat based on asymmetries observed in the phase-cycled bSSFP profile. The off-resonance encoding of steady-state magnetization, with phase-cycled bSSFP, may become a powerful alternative for fat fraction quantification, overcoming challenges of established time-domain encoding methods.

## 5. CONCLUSION

SPARCQ is a novel quantitative method for detection of different tissue components within the same voxel. This proof-of-concept study demonstrated the potential for voxel-wise water-fat separation and quantification by using asymmetries observed in the phase-cycled bSSFP signal profile. SPARCQ showed noise robustness, accuracy, and repeatability corroborated by simulations, phantom and volunteer experiments, and may be a promising alternative to conventional proton density fat fraction estimation techniques.

## 6. ACKNOWLEDGEMENTS

This study was supported by funding received from the Swiss National Science Foundation, grant PCEFP2_194296, the Emma Muschamp foundation, and the Swiss Heart foundation, grant FF18054. The authors acknowledge the use of the ISMRM Fat-Water Toolbox[15] for analysis of multiecho GRE data (http://ismrm.org/workshops/FatWater12/data.htm). The fat phantom was constructed thanks to access to the facilities and expertise of the CIBM Center for Biomedical Imaging, founded and supported by Lausanne University Hospital (CHUV), University of Lausanne (UNIL), École Polytechnique Fédérale de Lausanne (EPFL), University of Geneva (UNIGE) and Geneva University Hospitals (HUG).

## 7. DISCLOSURE

Several authors are inventors of a patent related to the proposed development[56]. Tom Hilbert and Tobias Kober are employed by Siemens Healthineers International AG.

## 8. DATA AVAILABILITY STATEMENT

The acquired data from phantoms and volunteers is publicly available and can be found on the following repository: https://zenodo.org/record/7789414#.ZCb5uXZBzcs.

# 9. SUPPORTING INFORMATION FIGURE CAPTIONS

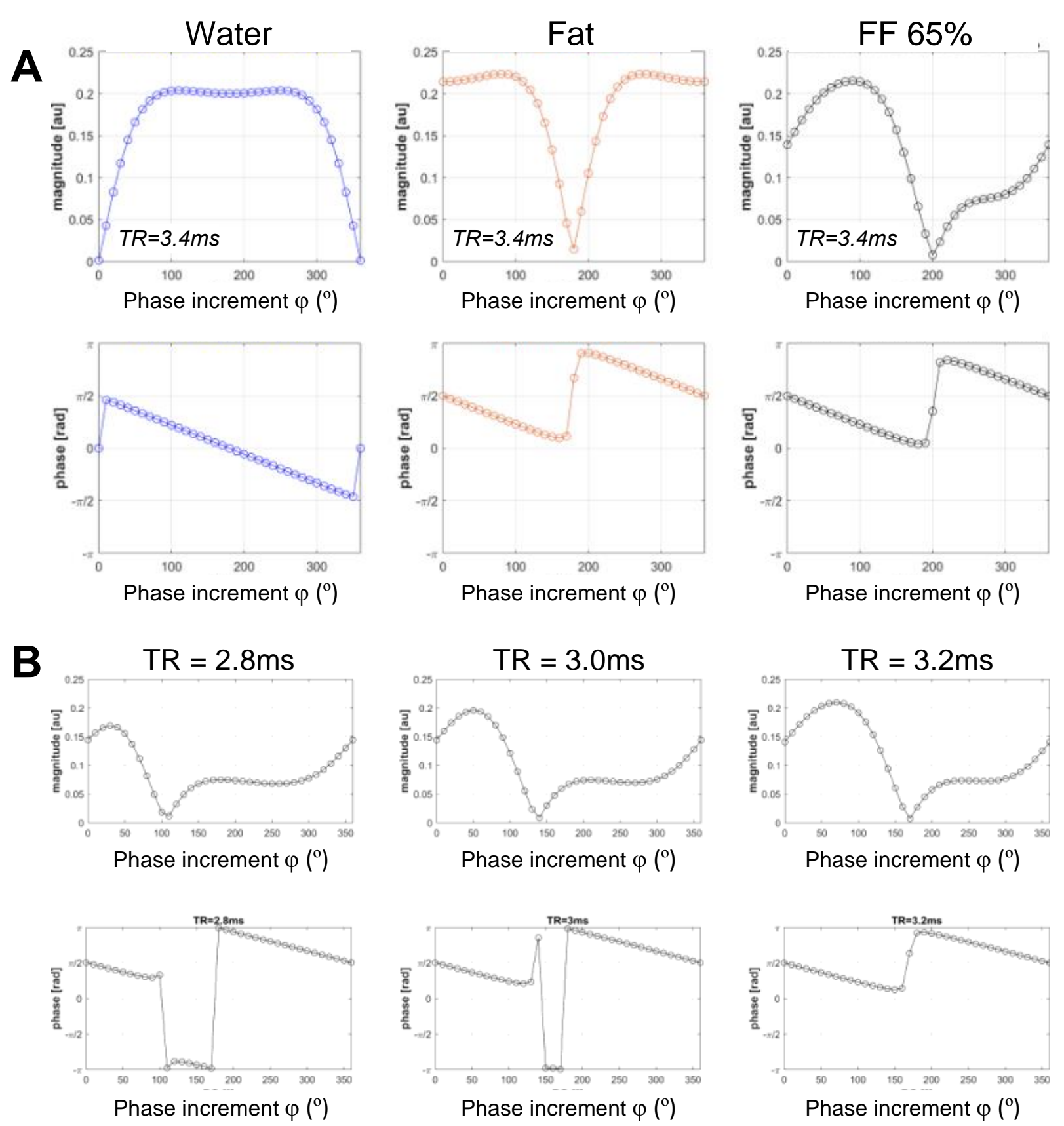

**Supporting Information Figure S1.**

(A) Simulated phase-cycled bSSFP profiles for a TR of 3.4ms for water (blue), fat (red), and a 65% fat fraction compartment (black). Notice that for a TR of 3.4 ms, the water and fat profiles are shifted so that their relative distances in the wrapped spectrum, in the range [0 1/TR], are maximized. (B) Simulated profiles for different values of TR for a 65% fat fraction compartment. Reducing the TR demonstrates that asymmetries are still present but change in appearance.

**Supporting Information Methods S1. Phase augmentation**

**Methods**: A preliminary phase augmentation step is performed to guide the final matching process. Following the fingerprinting approach[48], the best fitting bSSFP profile is selected, which is defined as the signal in the dictionary for which the dot product between its normalized magnitude profile and the normalized magnitude profile of the acquired signal is maximal. The phase profile of the best-fitting simulated bSSFP profile is selected, and the median difference with the phase profile of the acquired bSSFP profile is computed. The value obtained represents a phase shift between the dictionary entry and the acquired bSSFP profile and is used to augment the phase of the acquired profile.

Phase augmentation is a rapid initial estimation that does not require high precision. To study the effect on fat fraction quantification using SPARCQ in the presence and in the absence of phase augmentation, four different experimental cases were simulated, using 10 iterations, with the other simulation parameters as described in the manuscript in **Section 2.**3.

Case 1: SPARCQ without the phase augmentation step. This simulation is equal to the simulation performed in the main manuscript, of which the results are obtained using 100 iterations (**Figure 3**).

Case 2: SPARCQ with the phase augmentation step.

Case 3: SPARCQ without the phase augmentation step but with an additional constant phase offset of pi/4 added to the simulated phase-cycled bSSFP profiles.

Case 4: SPARCQ with the phase augmentation step and with an additional constant phase offset of pi/4 added to the simulated phase-cycled bSSFP profiles.

**Results**: The phase augmentation step has negligible influence on fat fraction estimation (**Supporting Information Figure S2**). When comparing the mean error of Case 1 and Case 2, there is a minor underestimation of the estimated fat fraction for fractions above 0.8 when applying the phase augmentation step. For the smaller fat fraction region, smaller than 0.2, the results are nearly identical. A detrimental effect on estimated fat fraction can be seen when a phase-offset is added to the profiles without applying phase augmentation (Case 3). Furthermore, the application of phase-augmentation shows robustness to adding constant phase-offsets, as results are identical when comparing Case 2 with Case 4.

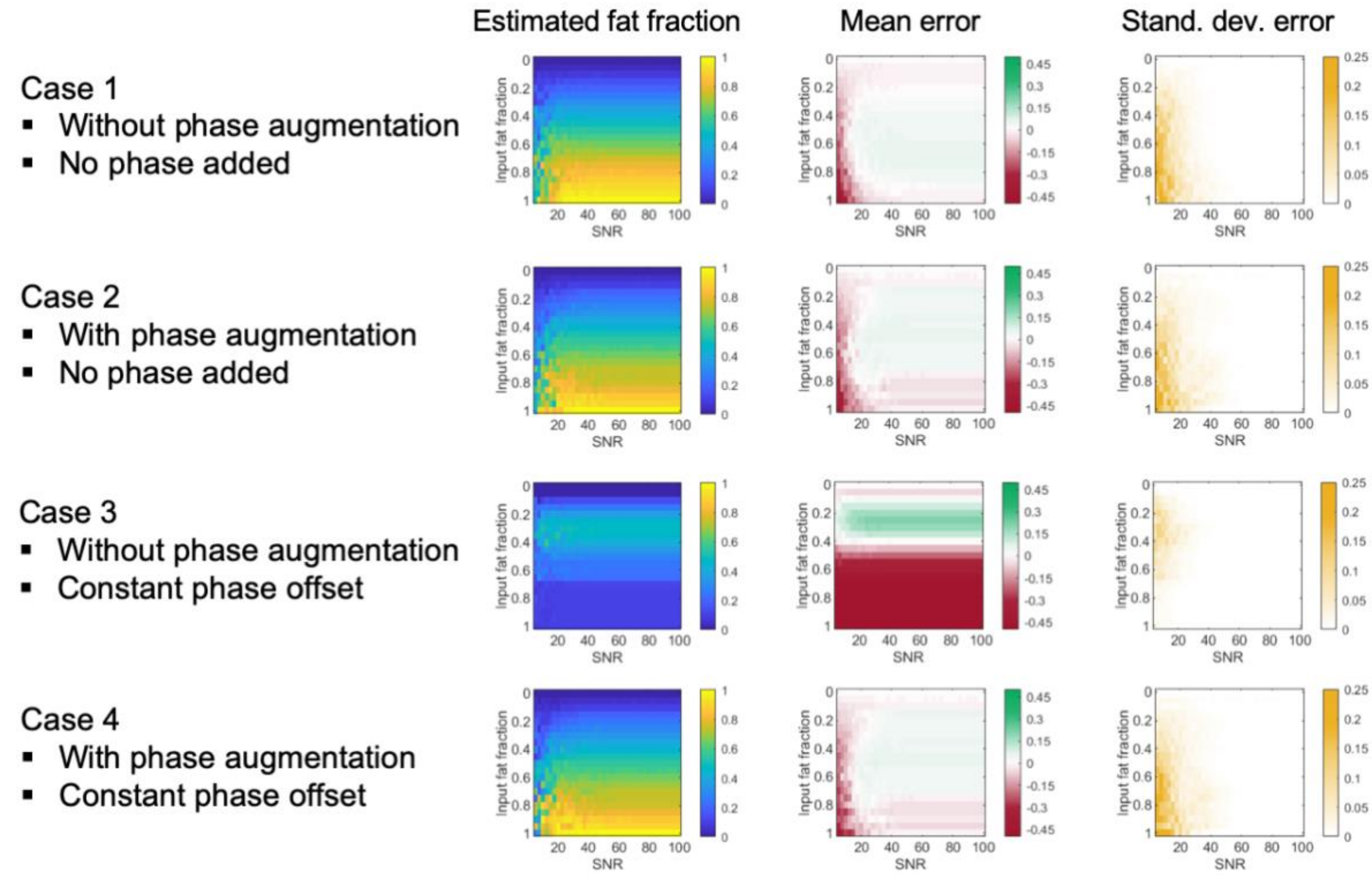

**Supporting Information Figure S2. Simulation experiments testing the effect of phase augmentation on fat fraction estimates.** Fat fractions were estimated with SPARCQ in four different cases. Case 1 simulated the base scenario that is similar to the simulations performed to obtain the results presented in **Figure 3**. Case 2 simulates the scenario where the phase-augmentation step is performed, which shows negligible effect on the estimated fat fraction. Case 3 demonstrates that when a constant phase-offset is added, the fat-fractions are severely over and underestimated in the absence of phase-augmentation. However, when phase augmentation is performed, the effects of adding a constant phase-offset are negated and fat-fraction estimates are nearly identical to Case 1 and Case 2.

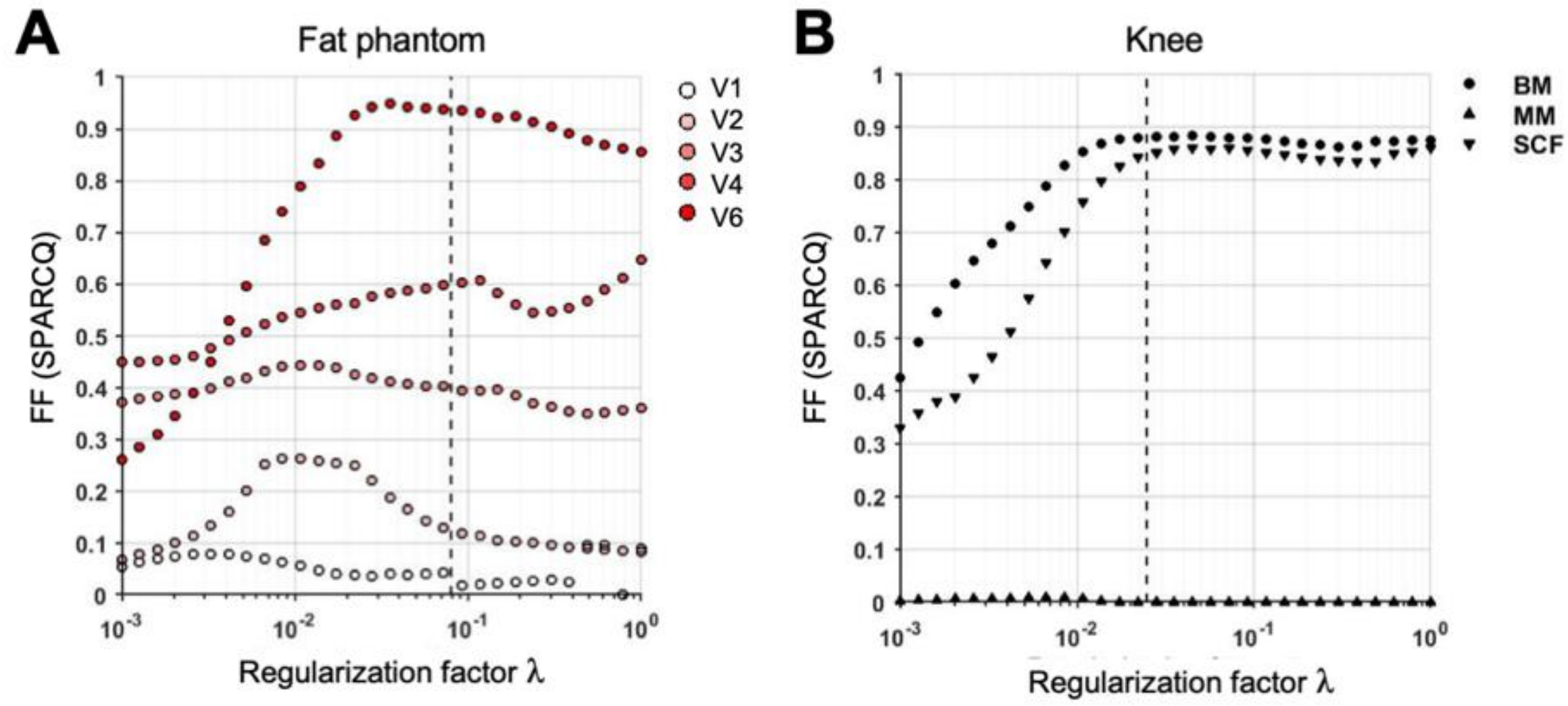

**Supporting Information Figure S3. Effect of regularization parameter on fat fraction estimations obtained with SPARCQ.** Fat fractions were estimated for logarithmically distributed values of the regularization parameter λ for five voxels in various phantom vials with increasing concentration (**A**, V80 was excluded due to phase separation) and for three voxels in three different knee tissues (**B**). Based on this manual adjustment, the lambda was set to 0.08 in phantom measurements and to 0.025 in volunteer experiments. BM : *bone marrow*, MM : *muscle*, SCF : *subcutaneous fat*.

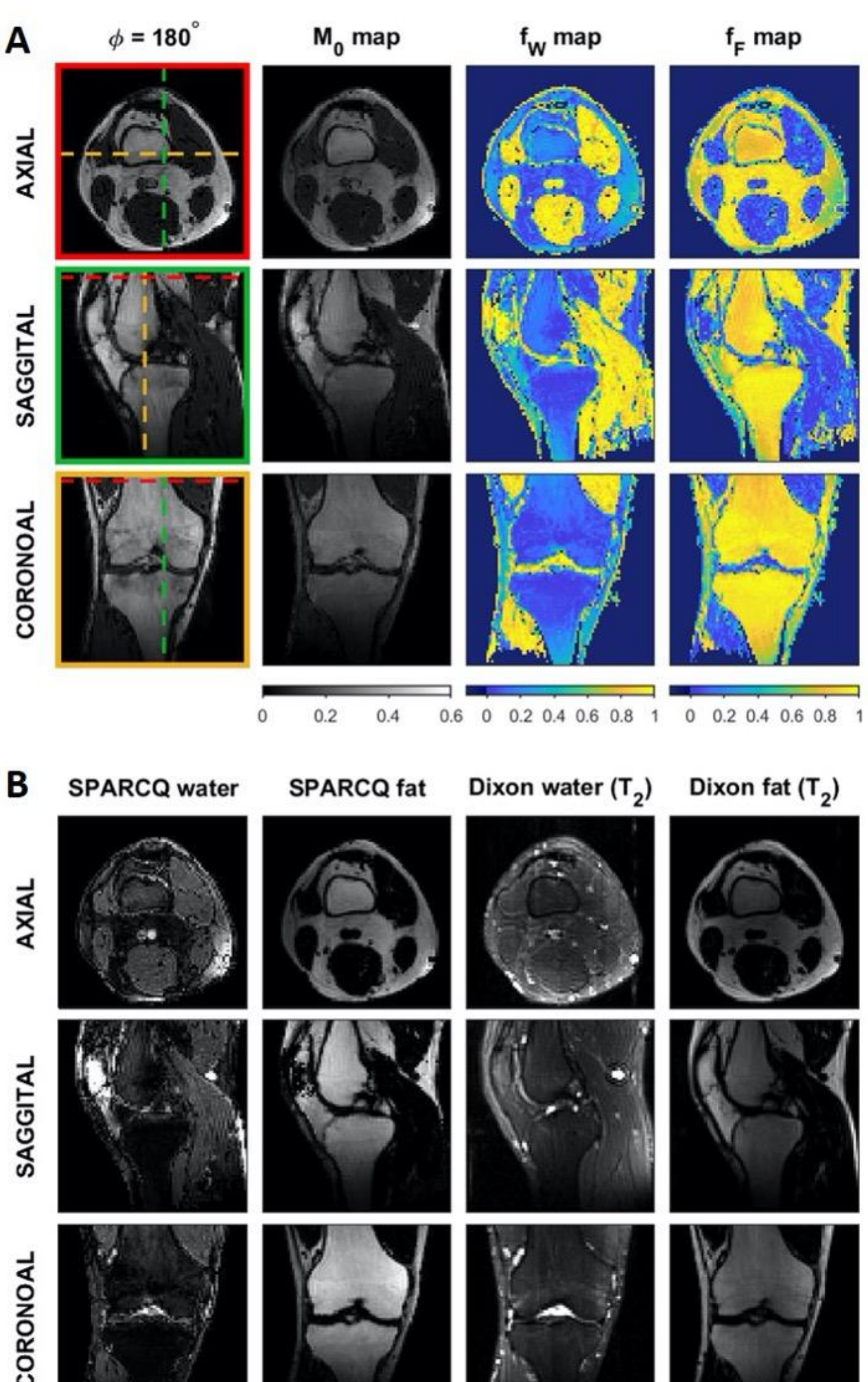

**Supporting Information Figure S4. Quantitative maps and water-fat-separated images.** Example images of one volunteer. (A) Quantitative maps obtained with SPARCQ in an axial, sagittal, and coronal view. (B) M0-weighted water-fat-separated images obtained with SPARCQ and T2-weighted water-fat-separated images obtained with TSE Dixon (TR/TE = 3470/100 ms, resolution $(2mm)^3$, receiver bandwidth 601 Hz/px, turbo factor 18; in-phase, out-of-phase, water and fat images were reconstructed on the scanner).

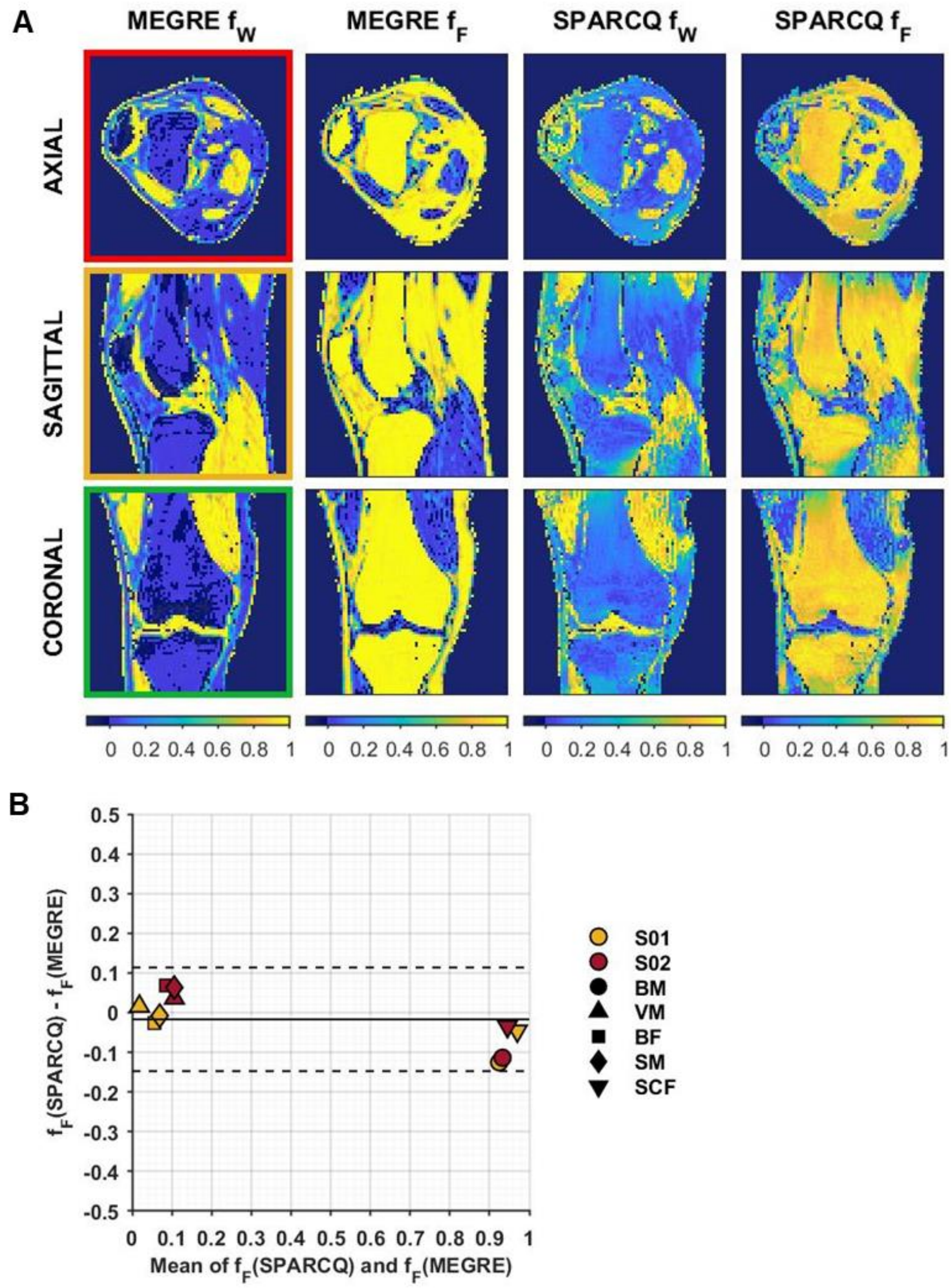

**Supporting Information Figure S5.** Quantitative water and fat fraction maps obtained with phase-cycled bSSFP (SPARCQ) and multiecho GRE (ME-GRE) in one volunteer. Bland-Altmann analysis of fat fractions estimated with SPARCQ and ME-GRE in 2 volunteers (1 volunteer = 1 color) for 5 regions of interest (1 marker type = 1 region). Limits of agreement : LOA = -0.0170 ± 1.96*0.0665. BM : *bone marrow*, VM : *vastus medialis*, BF : *biceps femoralis*, SM : *semimembranous muscle*, SCF : *subcutaneous fat*.

**Supporting Information Methods S2. $B_0$-induced errors in fat fraction estimates**

Although the current implementation of SPARCQ does not account for B0 inhomogeneities, these effects can be assessed with B0 maps derived from ME-GRE to reveal the induced error in fat fraction estimates. By not considering B0 inhomogeneities, the measured bSSFP profiles shift, resulting in estimation errors with SPARCQ as it uses fixed regions to assign water and fat. By performing experiments to determine the impact of $B_0$ inhomogeneities on SPARCQ-based fat fraction estimates, we can identify sources of error that can be mitigated in future implementations.

**Methods**: Phantom and volunteer data was used, where both SPARCQ and ME-GRE scans were performed. Based on the assumption that $B_0$ is constant throughout the duration of the scan, the effect of $B_0$ inhomogeneities on SPARCQ acquisitions was retrospectively mitigated using $B_0$ field maps obtained from ME-GRE acquisitions. First, voxel-wise $B_0$ field inhomogeneities (in Hz) were estimated from ME-GRE data using the ISMRM Fat-Water Toolbox. This information was used to compute the expected $B_0$ induced shifts in the bSSFP profiles. These shifts were used to circularly shift each complex bSSFP signal profile. The dictionary matching and calculation of the fat fraction maps were performed as previously described and compared with the maps obtained without shifting the bSSFP profiles according to the detected $B_0$. In phantom data, the obtained fat fractions in two slices close to air interfaces, therefore severely affected by $B_0$ inhomogeneities, were compared to the chemical shift thresholded fat fractions FF (CST) measured at 9.4T.

**Results**: In the phantom, $B_0$ correction effectively reduced estimation errors compared to the gold standard fat fractions (**Supporting Information Figure S6A**). In volunteers, in regions of high $B_0$ inhomogeneity the $B_0$-corrected fat fraction maps provided sensible improvement on fat fraction estimations (e.g. patella) with improved delineation of anatomical details (**Supporting Information Figure S6B**).

**Conclusion**: The experiment shows that $B_0$ inhomogeneities are responsible for erroneous fat fraction estimates (e.g. regions close to air-tissue interfaces in the knee) and that taking B0 inhomogeneities into account in the reconstruction process reduces the error with respect to gold-standard fat fractions (in the phantom). Embedding a future $B_0$ correction strategy within the SPARCQ framework may bring significant improvements.

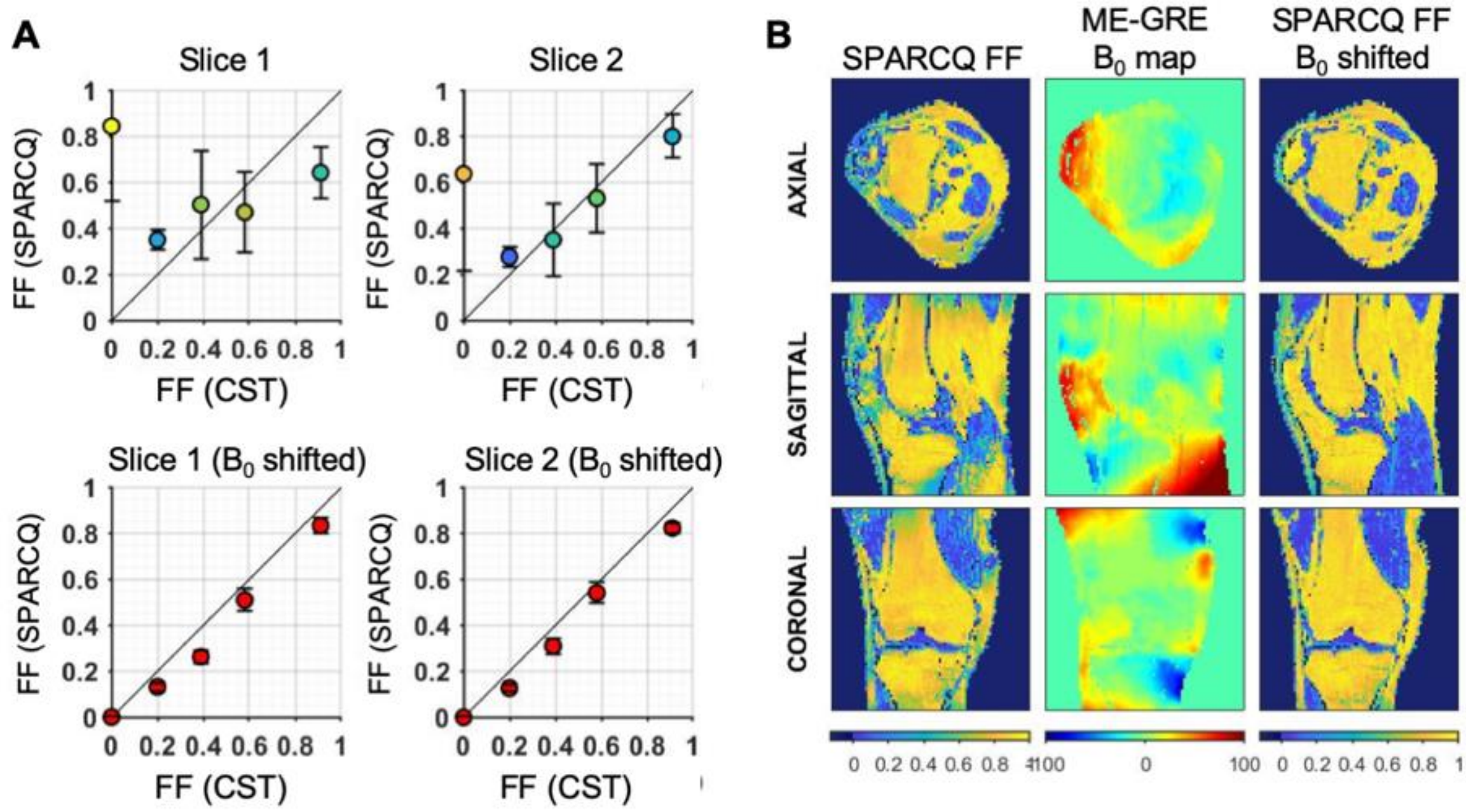

**Supporting Information Figure S6**. Quantitative graphs and maps with and without applied $B_0$ shift in different positions of the fat phantom (A) and different orientations in the knee of a volunteer (B).